# On the remote interaction of biological objects with identical genetic structures

Simon Berkovich
Dept. of CS
The George Washington University
Washington, DC 20052
USA
Phone:   (202)-994-8248
Fax:     (202)-994-4875
E-mail:  berkov@seas.gwu.edu

**Abstract**

The paper puts forward an unusual prediction that cultivating a clone can curtail the lifespan of the clone donor. Neither the arrangement of this suggested empirical study nor the analyses of the anticipated outcomes rely on the accompanying theoretical contemplations. This prediction has come from the interpretation of the genome as a "barcode". The genome is considered as an identification label rather than a repository of control information, so living beings are portrayed as a community of users on the "Internet of the physical Universe". Thus, biological objects with identical DNA structures can interfere, and the surmised remote impact appears tangible. The effect of clone-donor interaction leads to a decisive *Experimentum Crucis* that can reject the common view on the organization of biological information processing. Exploitation of this effect can be potentially dangerous.

**Keywords**:   biological information processing, genome, clones, organism development, longevity, antibiotic resistance

## 1. Introduction

Genome manipulations bring a feeling of unknown prospects. Newly created portions of DNA might be able to spread and produce unforeseen consequences in the long run. This paper suggests a more dramatic scenario where genetic manipulations may cause a direct impact through informational cross-interference. Namely, it is suspected that cultivating a clone can curtail the lifespan of the clone donor.

Investigations of the genome have discovered several confusing circumstances. First, the amount of information that is contained in 30,000 genes of humans is very low in an absolute sense, it is less than the amount of information in a blurry picture taken by a substandard digital camera. Second, humans have less genetic information in relative sense with respect to apparently simpler biological objects, like rice, which contains 32,000-55,000 genes. Third, nearly all (about 97%) of the genome - "junk DNA" - does not seem to contribute to biochemical workings of the cell. From the standpoint of information theory, considering the genome that can hardly store a blurry view of a living creature as a "blueprint" for organism development is naïve and inept.

Purposeful behavior requires a continuous influx of control information. Thus, in the first place, the phenomenon of Life is very intensive information processing. Biological information processing must comply with "the basic law of requisite variety", which says that achieving of appropriate selection "is absolutely dependent on the processing of at least that quantity of information. Future work must respect this law, or be marked as futile even before it has been started" (Ashby, 1962). According to our suggestion (Berkovich, 1999 and 2001), the DNA is not a repository of control information but an identification key for an organism, like a "barcode". Physico-chemical processes are too slow and unreliable to implement robust high-speed information processing for the control of living systems. With the DNA as a key, biological objects have access to richer information processing facilities of the underlying infrastructure of the physical world. Thus, the length and composition of DNA structures are not related to the complexity of organisms they represent. This observation is well documented but basically ignored.

The paper confronts the existing view on biological information processing with the hypothesis on remote interaction of biological objects having close DNA structures. It should be emphasized that empirical investigations of the suggested hypothesis do not require any specific theoretical backup. Yet the scientific implications of this hypothesis are overwhelming. The corresponding theoretical constructions are outlined elsewhere (see, in particular, Berkovich (2001) and numerous references therein). It might not be inspiring to get into all these problems unless the suspected effect of remote biological impact actually materializes.

As a primary test of the surmised interference of biological objects with identical DNA structures, Section 2 discusses the possibility of shortening of the lifespan of clone donors due to cultivating of their clones. Some notes in Section 3 provide an indirect support to this anticipation. Concluding commentaries are given in Section 4.

## 2. The hypothesis on shortened longevity of clone donors

The rationale for the hypothesis that cultivating clones can shorten the lifespan of clone donors is given below. For the empirical utilization of this hypothesis the presented theoretical speculations are not essential.

A zygote starting a new organism opens "an account" on the "Internet of the physical Universe" with the DNA as a key. One part of this account is used in read-only mode for the construction of the organism; the other part is used for write-once-read-many mode for operational control. Different biological objects get distinct "slices" of the information processing resources of the Universe corresponding to their DNA determined "accounts". Biological objects share the storage and bandwidth of the informational infrastructure of the physical world in the Code Division Multiple Access (CDMA) mode, similarly to cellular phone communications.

The "Internet-like" organization with the underlying infrastructure of the physical world raises a particular concern how zillions of biological objects avoid informational cross-interference. The feasibility of this construction, which employs the CDMA communications, hinges upon biological individuality of multi-cellular organisms - the distinctiveness of the DNA keys. Multi-cellular organisms acquire this property through randomization in meiosis when pieces of paternal chromosomes undergo a chaotic recombination. In the decentralized randomized assignment, chances that the DNA keys would come out alike are very low. Such an undesirable possibility is a malfunction of the system. It results in informational cross-talks of multi-cellular organisms, which, for example, might show up as extra-sensory perception. However, since appearances of this system malfunction are rare and sporadic, observations of extra-sensory perception are non-reproducible and inconclusive.

On the other hand, in the case of clones, biological objects having exactly matching DNA are generated artificially. Thus, nowadays, due to new facilities provided by cloning technology, the informational cross-talks of biological objects can be investigated in repetitive controlled conditions. A newly produced clone, in contrast to a sexually created organism, simply takes the existing DNA key and enters the already established "account" of the donor. Not surprisingly, a clone inherits the "informational" age of the donor and undergoes premature aging and early death. Clones often fail to develop into normal organisms. But this abnormal development is not because their genetic structure is distorted. The genetic structure of clones would be presumed in a good shape as long as the clones can produce healthy offsprings. Indeed, the development of a clone is based on a partially used up "account" of the donor, while the development of a clone offspring, which gets a re-hashed DNA key, starts afresh.

Each account has a finite life cycle due to limited capacity to efficiently overcome the increase of accumulated information. A clone and its donor, which have a joint "account" each, handle more information, so the operational capacity of this account for each user will be exhausted in a shorter period of time.

An operational step of biological control is associated with inputting a certain portion of information into the organism's "account". Let us assume that inputs units of a clone and its donor have a common part specified by an overlap factor, ρ (0 ≤ ρ ≤ 1). Because of this overlapping, a clone and its donor will jointly input into their account (2 – ρ) units of information rather than 2 units of information if these units of information were completely different. So, the remaining part of donor's life as a result of a clone joining this account will be cut by a factor (2 – ρ). The longevity of the clone itself will be equal to the unused remainder of life of the donor divided by (2 – ρ). In the case ρ = 0 (no overlap) the longevity of the clone will be half of the unused remainder of donor's life, in the case ρ = 1 (complete overlap) the longevity of the clone will be the same as the unused remainder of the donor's life. The leftover duration of the donor's life will be changed correspondingly.

Now, consider a group of N clones from the same donor. Let us also assume that inputs of these (N+1) objects overlap pairwise on a common part ρ and further on in a random fashion, so triplet overlappings occupy $\rho^2$ part of the input unit, quadruplet overlappings occupy $\rho^3$, and so on. Applying the principle of inclusion and exclusion the union of input units, U, of all (N+1) objects – the donor and N clones – can be represented as:

$$U = (N+1) - \binom{N+1}{2} \cdot \rho + \binom{N+1}{3} \cdot \rho^2 - \binom{N+1}{4} \cdot \rho^3 + \ldots = [1 - (1-\rho)^{N+1}]/\rho \qquad (1)$$

Thus, in a group of a donor and N clones the relative value, L, presenting the leftover lifespan of the donor and the longevities of the clones as a fraction of the remainder of donor's life, will be the inverse of the expression (1):

$$L = \rho/[1 - (1-\rho)^{N+1}] \qquad (2)$$

In the case ρ = 1 (complete overlap), L = 1 – the lifespans of the donor and the clones are not affected. In the case ρ → 0 (no overlap), L → 1/(N+1) – the lifespan of each object is cut by the total number of the objects using the same account.

For a one-clone system the remaining part of donor's life is 1/(2 – ρ). When the number of clones increases the remaining part of donor's life tends to ρ. Evidently, donors lifespans are shortened more when the overlap of clones inputs, ρ, is small.

Recent experiments indicate that cloned animals in fact have a shorter lifespan (Cohen, 2002). As long as this result is applied to clones only it can be furnished with many different explanations not related to mutual interference. *The longevity of clone donors, presumably, has not been examined.* What could be expected for the longevity of clone donors can be roughly estimated from the available data provided that the hypothesis of mutual interference is correct.

In the experiments reported in (Cohen, 2002) 10 out 12 cloned mice were dead after 800 days; for the control animals all but 3 of the 13 were still alive. In the absence of better reasons, we can use linear extrapolation to project the regular lifespan of the control animals. So, a conservative estimate for the relation of this regular lifespan to the lifespan of the clones might be about 3. Thus, we take $\rho = 0.3$. This means that for a many-clones system the lifespan of a donor should constitute about 30% of a regular remaining duration of its life. From this we derive that in one-clone system the lifespan of a donor will be shortened only to $1/(2 - 0.3) = 0.6$, i.e. the lifespan of a donor with one clone will constitute about 60% of its remaining life. So, if the hypothetical effect of shortening the lifespan of clone donors actually takes place, its outcome would be quite observable.

## 3. Related circumstances

The first controlled study of the lifespan of clones (Cohen, 2002) showed that a batch of cloned mice all died earlier than their naturally bred cousins. This fact can be furnished with various explanations, but it definitely clears the suggested hypothesis on possible cross-interference of biological objects with close DNA. In other words, if the effect of shortening of clones lifespans would not be observed, our hypothesis would be questioned. Furthermore, the lifespans of clones should be inversely dependent on the number clones in the group produced from a given donor. These observations as a by-product of regular investigations of clones can represent an easygoing version of the suggested study of mutual influences of clones and donors.

The idea of cross-interference of biological objects with close DNA is supported by a number of indirect evidences. Thus, restricting the extent of biological activities reduces the intensity of information control transactions and as a consequence spares the organism's "account". This agrees with an established fact that limitations in food consumption can noticeably increase the longevity of an organism. The longevity is also affected by the size of biological objects: *ceteris paribus*, smaller organisms of the same species live longer. Besides cloning, biological objects with close DNA structures to a certain degree occur in sexual reproduction. Therefore, information interference of the near kin may be involved in correlation of the longevity with the performance of the reproduction system (Hsin and Kenyon, 1999, Westendorp and Kirkwood, 1998).

Informational cross-interference of organisms with closely matching DNA might be most effectual for monozygotic twins, which appear similar to clones in this respect. The control information for organism development and for operational procedures may be handled by different subsets of chromosomes. It was found out that the probability of monozygotic twins is an integer power of ½, and this property has been associated with non-advancement of the cell differentiation counter at zygote division (Berkovich and Bloom, 1985). For humans, the probability of monozygotic twins is $1/256 = (½)^8$ suggesting that 8 pairs of chromosomes are involved in regulating organism development. In acquiring their biological individuality monozygotic twins can rely only on the rest of the 15 pairs of chromosomes. Thus, genetic similarities, and hence cross-talk capabilities, among twins are higher than among siblings but lower than among clones.

In single cell organisms, like bacteria, the DNA structures are generated by mere replications, so they are identical. Cross-talks among such biological objects are more pronounced and populations of bacteria can create interrelated communicating networks (Sonea, 1988). These networks are fixed to the same "account" since new bacteria get the same DNA keys as their progenitors. A particular "account" corresponding to a given DNA key has limited operational capabilities. Therefore, to survive populations of microorganisms have to undergo continuous transformations, so that altered DNA keys would enable switching from one used up "account" to another. The emergent resistance of bacteria to antibiotics may reflect workings of such a mechanism. Remarkably, observations (Gilliver et al., 1999) show a buildup of resistance to antibiotics in the absence of a traceable exposure. Thus, it is important to examine whether germination of drug resistant pathogenic microorganisms in a firmly isolated localized environment may lead to a proliferation of such microorganisms on a global scale.

**4. Conclusion**

The most valuable instrument for the advancement of knowledge is an *Experimentum Crucis* - a crucial experiment capable to reject an existing theory. A good crucial experiment is very rare since unusual outcomes are readily challenged by alternative explanations. The investigation of the longevity of clones has an unprecedented convincing power – modern science just cannot envision a situation when the lifespan of a living being can be curtailed from a remote location without a tangible material contact.

The idea that Life and Mind involve activities beyond the ponderable matter had flourished in the XIX century with the development of aether theory. In widely acclaimed book of 1873, *The Unseen Universe*, Stewart and Tait wrote: "We attempt to show that we are absolutely driven by scientific principles to acknowledge the existence of an Unseen Universe, and by scientific analogy to conclude that it is full of life and intelligence - that it is in fact a spiritual universe and not a dead one" (cited from (Powers, 1985)). In the XX century, with elevation of relativity the idea of aether has been abandoned, although "undeservedly" as regarded by Wilczek (1999). The concept of relativity lends to two interpretations: according to Einstein the absolute frame of reference does not exist, according to Lorentz and Poincaré the absolute frame of reference is simply undetectable. For practicing physics adhering to either of these interpretations is inconsequential. As to the biology, the later interpretation may expose attributes of the infrastructure of the physical world for information processing.

Reduced to its essence, the concept of relativity imposes a postulate that everything in the world is ultimately determined only by configurations of the material structures involved. Rejecting this postulate in conjunction with the phenomenon of Life brings up the idea of extracorporeal organization of biological information processing. Living beings get control information from the infrastructure of the material world creating an Internet-like community of users. Figuratively speaking, human brain is a terminal at the "Internet of the physical Universe" rather than a stand-alone computer (Berkovich, 1993). From the viewpoint of the contemporary methodology of information systems design the Internet-like organization of biological information processing makes a lot of sense.

One of the striking experiences revealing the separation of biological information processing from the material substrate comes from observations of "near death experience", which is a reported recollection of impressions during a special state of consciousness, including such elements as out-of-body experience, pleasant feelings, and seeing a tunnel, a light, deceased relatives, or a life review (Lommel, 2001). These observations lead to a great commotion: "How could a clear consciousness outside one's body be experienced at the moment that the brain no longer functions during a period of clinical death with flat EEG?"

Amidst theoretical controversies surrounding the idea of extracorporeal organization of biological information processing, it would be imprudent to ignore the pragmatic perspective of this organization. By elaborating various cloning techniques "biological hackers" on the "Internet of the physical Universe" can perpetrate a misconduct of "stolen identity". Thus, developing clones from DNA samples of a target organism can destroy this organism remotely. Following our rough preliminary estimates, cultivating a multiplicity of clones may cut the remaining part of donor's life by about 3 times.

Shortening of the longevity of clone donors is a simple empirical circumstance that can be discovered and comprehended independently of the involved theoretical conceptions. For different species the effect of remote biological impact develops in different time scales. For example, it would take several months to observe the occurrence of shortened longevity for mice. As to humans, this observation may take several years. So, remote biological impact can become operational before the meaning of this situation is fully realized.

## 5. Acknowledgement